\documentclass[aip,jcp,twocolumn,reprint]{revtex4-1}

\usepackage{color}
\usepackage{graphicx}
\usepackage{tabularx}
\usepackage{amssymb}
\usepackage{amsmath}
\usepackage{bm}
\usepackage{graphpap}
\usepackage{multirow}
\usepackage{notes2bib}
\usepackage[version=3]{mhchem} % Formula subscripts using \ce{}

\begin{document}
%%%%%%%%%%%%%%%%%%%%%%%%%%%%%%%%%%%%%
\title{Optical properties of azobenzene-functionalized self-assembled monolayers: Intermolecular coupling and many-body interactions}
%%%%%%%%%%%%%%%%%%%%%%%%%%%%%%%%%%%%%%%%%%%%%%%%%%%%%%%%%%%%%%%%%%%%%
\author{Caterina \surname{Cocchi}}
\email{caterina.cocchi@physik.hu-berlin.de}
\affiliation{Institut f\"ur Physik and IRIS Adlershof, Humboldt-Universit\"at zu Berlin, Berlin, Germany}
\affiliation{European Theoretical Spectroscopic Facility (ETSF)}
\author{Thomas \surname{Moldt}}
\affiliation{Fachbereich Physik, Freie Universit\"at Berlin, Berlin, Germany}
\author{Cornelius \surname{Gahl}}
\affiliation{Fachbereich Physik, Freie Universit\"at Berlin, Berlin, Germany}
\author{Martin \surname{Weinelt}}
\affiliation{Fachbereich Physik, Freie Universit\"at Berlin, Berlin, Germany}
\author{Claudia \surname{Draxl}}
\affiliation{Institut f\"ur Physik and IRIS Adlershof, Humboldt-Universit\"at zu Berlin, Berlin, Germany}
\affiliation{European Theoretical Spectroscopic Facility (ETSF)}

%%%%%%%%%%%%%%%%%%%%%%%%%%%%%%%%%%%%%%%%%%%%%%%%%%%%%%%%%%%%%%%%%%%%%
\begin{abstract}
In a joint theoretical and experimental work the optical properties of azobenzene-functionalized self-assembled monolayers (SAMs) are studied at different molecular packing densities.
Our results, based on density-functional and many-body perturbation theory, as well as on differential reflectance (DR) spectroscopy, shed light on the microscopic mechanisms ruling photo-absorption in these systems.
While the optical excitations are intrinsically excitonic in nature, regardless of the molecular concentration, in densely-packed SAMs intermolecular coupling and local-field effects are responsible for a sizable weakening of the exciton binding strength.
Through a detailed analysis of the character of the electron-hole pairs, we show that distinct excitations involved in the photo-isomerization at low molecular concentrations are dramatically broadened by intermolecular interactions.
Spectral shifts in the calculated DR spectra are in good agreement with the experimental results.
Our findings represent an important step forward to rationalize the excited-state properties of these complex materials.
\end{abstract}
%%%%%%%%%%%%%%%%%%%%%%%%%%%%%%%%%%%%%%%%%%%%%%%%%%%%%%%%%%%%%%%%%%%%%

\date{\today}
%
%%%%%%%%%%%%%%%%%%%%%%%%%%%%%%%%%%%%%%%%%%%%%%%%%%%%%%%%%%%%%%%%%%%%%
\maketitle
%*******************************
%  INTRODUCTION
%*******************************
\section{Introduction}
Controlled architectures of photo-responsive chromophores can be obtained in an elegant way by self-assembly of functionalized molecules in ordered monolayers.
In this way, photo-switching materials can be potentially exploited in view of realistic applications. \cite{brow-feri06natn,balz+08chpch,russ-hech10am,fuen+11ns,tege12jpcm,aben+15nano}
Azobenzene-functionalized self-assembled monolayers (SAMs) of alkyl chains have been successfully synthesized in the last two decades. \cite{wolf+95jpc,zhan+97cpl,yasu+03jacs,delo+05lang,kuma+08nl,klaj10pac}
Unfortunately, such systems come with a substantial drawback: The photo-isomerization rate is drastically suppressed due to steric hindrance or excitonic coupling. \cite{gahl+10jacs,vall+13lang}
A number of experimental strategies have been suggested to overcome this problem, such as replacing the aliphatic linker by an aromatic one \cite{pace+07pnas} or even bulkier groups, \cite{jung+10lang,jung+12jpcc,jaco+14pccp} or diluting the azobenzene moieties in densely-packed alkanethiolate SAMs. \cite{vall+13lang,evan+98lang,naga+09jesrp,mold+15lang,mold+16lang}

From a theoretical viewpoint the issue of hindered photo-isomerization in densely-packed azobenzene-functionalized SAMs has been addressed by a few works, mainly focused on the excitonic coupling between the chromophores, \cite{utec+11pccp,bena-corn13jpcc} as well as on the effects of a metal substrate. \cite{bena-corn14jpcc}
While these investigations have significantly contributed to describe the excited-state properties of azobenzene derivatives, a very recent publication elucidates the role of defects in the photo-isomerization of such SAMs. \cite{cant+16jpcl}
Still, a full understanding of the fundamental physical mechanisms ruling optical absorption at increasing molecular concentration is still missing. 
Identifying the nature of the excitations is an essential step in view of defining new strategies that enable to restore the photo-switching efficiency exhibited by azobenzene in solution.
For this purpose, we provide here an in-depth analysis of the basic physical mechanisms governing photo-absorption in well-ordered azobenzene-functionalized SAMs at increasing packing density of the chromophores.
To do so, we combine \textit{ab initio} many-body theory with differential reflectance spectroscopy.
We investigate the effects of increasing molecular density on the photo-absorption properties of these systems, specifically focusing on the role of intermolecular coupling and local-field effects.
The analysis of the spectra is supported by a detailed characterization of the optical excitations, in view of explaining how many-body effects rule the excitation process.
Good agreement between theory and experiment corroborates our conclusions.

%
%*******************************
%  SYSTEMS
%*******************************
\section{Self-assembled monolayers of azobenzene-functionalized alkanethiols}
\label{sec:systems}
\begin{figure}%[h!]
\centering
\includegraphics[width=.48\textwidth]{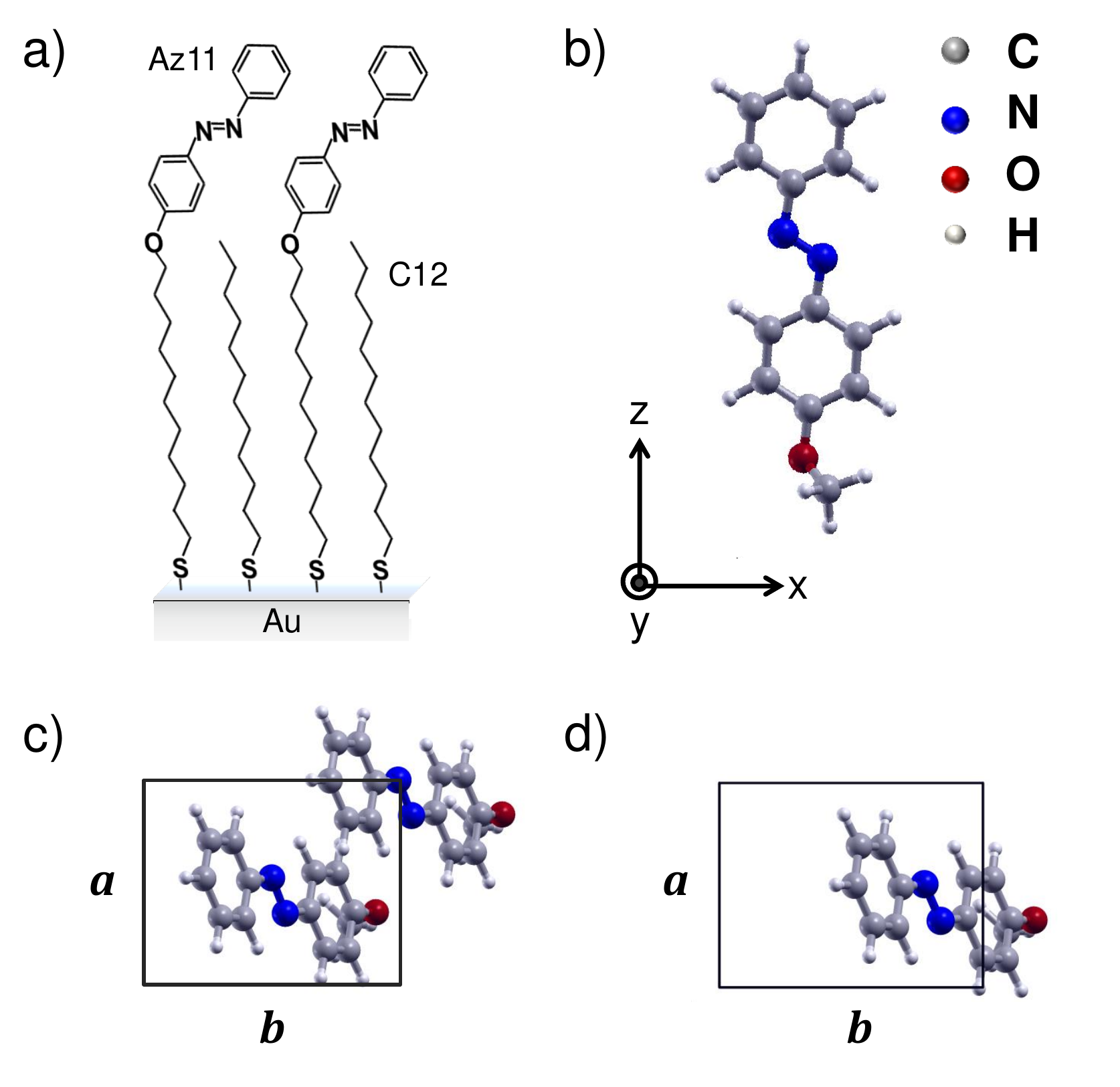}%
\caption{(Color online) a) Sketch of the experimentally investigated azobenzene-functionalized SAM of alkanethiols on a gold surface. b) Ball-and-stick representation of an azobenzene molecule with \ce{OCH3} termination. Unit cell adopted in the first-principles calculations to model c)
a packed azobenzene-functionalized SAM (p-SAM) and d) its diluted counterpart (d-SAM).
}
\label{fig:H-az}
\end{figure}
%%%%%%%%%%%%%%%%%%%%%%%%%%%%%%%%%%%%%%%
% 
In this work, we focus on azobenzene-functionalized alkanethiolate SAMs, which form well-ordered, densely-packed structures with the \textit{trans}-configuration representing the ground state. 
Intermolecular interactions in such systems result in a predominantly upright orientation of the alkyl chains as well as of the chromophore (see Fig.~\ref{fig:H-az}a). 
On average, the plane of the azobenzene moiety is tilted by 73$^{\circ}$ with respect to the surface. \cite{gahl+10jacs}
The lateral SAM structure has been investigated by atomic-force microscopy (AFM) and scanning tunneling microscopy (STM). \cite{jasc+96jpc,mann+02jpcb}
Densely-packed azobenzene-alkanethiolate SAMs form a nearly rectangular two-dimensional structure with lattice constants $a$=6.05\,\AA{} and $b$=7.80\,\AA{} including two molecules in the unit cell. 
Without sacrifying the overall structure of the SAM, the density of azobenzene moieties can be reduced by mixing functionalized and unfunctionalized alkanethiolates. As a consequence of the weaker interaction between the chromophores photo-isomerization is enabled. Additionally, the azobenzene moieties tend to orient more parallel to the surface. For a dilution up to 20$\%$ of a densely-packed azobenzene SAM the average tilt angle of the aromatic plane is about 45$^{\circ}$. \cite{mold+15lang}
%

%%%%%%%%%%%%%%%%%%%%%%%%%%%%%%%%%%%%%%%
%\subsection{Theoretical Modeling}

In our first-principles calculations, azobenzene-functionalized SAMs are modeled  by a two-dimensional crystal structure, where the alkyl chains and the gold surface are not considered.
This choice is motivated by the fact that the alkyl chains do not contribute significantly to the optical absorption at photon energies below 6\,eV, and that the azobenzene moieties are expected to be decoupled from the gold substrate by the alkyl chains.
To mimic the chemical environment of the bond to the alkyl chain, we add a methoxy (\ce{OCH3}) group terminating the azobenzene molecules (see Fig.~\ref{fig:H-az}b).
We model a densely-packed SAM (p-SAM) by considering an orthorhombic unit cell \bibnote{For the reciprocal-space representation of an orthorhombic unit cell and the identification of the corresponding high-symmetry points, we refer to the Bilbao Crystallographic Server. \cite{aroy+11bcc}} of lattice parameters $a$ and $b$, incorporating two molecules oriented parallel to each other and tilted by 30$^{\circ}$ with respect to the surface normal corresponding to the $ab$ plane (Fig.~\ref{fig:H-az}c).
In this configuration the chromophores are separated by $\sim$2\,\AA{} and $\sim$3.8\,\AA{} in the direction parallel and perpendicular to the plane of the phenyl rings, respectively.
We include $\sim$14\,\AA{} of vacuum in the normal direction, to avoid unphysical interactions between the replicas.
To model a SAM with reduced packing density, we consider a system including only one molecule per identical unit cell.
In the following, we refer to the structure shown in Fig.~\ref{fig:H-az}d as diluted SAM (d-SAM).
For comparison with the isolated molecule, we study also a single azobenzene chromophore in an orthorhombic supercell, incorporating $\sim$6\,\AA{} of vacuum in each direction.
The electronic and optical properties of the investigated systems depend essentially on intermolecular distances. 
Only minor effects are expected to come from the orientation of the molecules in the unit cell and with respect to each other. 
For this reason, we can rely on the geometries adopted in our first-principles calculations, although rather simplified compared to the experimental sample.

%%%%%%%%%%%%%%%%%%%%%%%%%%%%%%%%%%%
%*******************************
%  METHODS
%*******************************
\section{Theoretical background }

Ground-state properties are obtained in the framework of density-functional theory (DFT), adopting the linearized augmented planewave (LAPW) plus local orbital method as implemented in the \texttt{exciting} code. \cite{gula+14jpcm}
Optical absorption spectra are calculated from many-body perturbation theory, following a two-step procedure: Quasi-particle (QP) energies are evaluated from $G_0W_0$ \cite{hedi65pr,hybe-loui85prl} and optical excitations are computed with the Bethe-Salpeter equation (BSE), an effective equation of motion for the electron-hole ($e$-$h$) two-particle Green's function. \cite{hank-sham80prb,stri88rnc}
This methodology allows for describing optical excitations in molecular materials from the gas-phase to the condensed matter. \cite{hiro+15prb,jacq+15jctc,cocc-drax15prb1,brun+15jcp,ruin+02prl,pusc-ambr02prl,humm+04prl,humm-ambr05prb}
In matrix form, the BSE is expressed as
\begin{equation}
\sum_{v'c'\mathbf{k'}} \hat{H}^{BSE}_{vc\mathbf{k},v'c'\mathbf{k'}} A^{\lambda}_{v'c'\mathbf{k'}} = E^{\lambda} A^{\lambda}_{vc\mathbf{k}} ,
\label{eq:BSE}
\end{equation}
where the indexes $v$ and $c$ indicate valence and conduction states, respectively.
In a spin-unpolarized system the effective two-particle Hamiltonian \cite{rohl-loui00prb,pusc-ambr02prb} reads:
\begin{equation}
\hat{H}^{BSE} = \hat{H}^{diag} + 2 \gamma_x \hat{H}^x + \gamma_c \hat{H}^{dir}.
\label{eq:H_BSE}
\end{equation}
$\hat{H}^{diag}$ is the \textit{diagonal} term, which accounts for single-particle transition energies.
Considering only this term in Eq.~\ref{eq:H_BSE} corresponds to the independent-particle approximation (IPA).
The term $\hat{H}^x$ includes the short-range exchange Coulomb interaction $\bar{v}$, which accounts for local-field effects (LFE):
\begin{equation}
\hat{H}^x = \int d^3\mathbf{r} \int d^3\mathbf{r}' \phi_{v\mathbf{k}} (\mathbf{r}) \phi^*_{c\mathbf{k}} (\mathbf{r}) \bar{v}(\mathbf{r},\mathbf{r}') \phi^*_{v'\mathbf{k}'} (\mathbf{r}') \phi_{c'\mathbf{k}'} (\mathbf{r}'),
\label{eq:Hx}
\end{equation}
where $\phi$ are QP wave-functions.
$\hat{H}^{dir}$ contains the statically screened Coulomb potential $W$, which describes the attractive $e$-$h$ interaction:
\begin{equation}
\hat{H}^{dir} \! \! = \! - \! \! \int \! \! \! d^3\mathbf{r} \! \!  \int \! \! d^3\mathbf{r}' \phi_{v\mathbf{k}} (\mathbf{r}) \phi^*_{c\mathbf{k}} (\mathbf{r}') W(\mathbf{r},\mathbf{r}') \phi^*_{v'\mathbf{k}'} (\mathbf{r}) \phi_{c'\mathbf{k}'} (\mathbf{r}').
\label{eq:Hdir}
\end{equation}
The screened Coulomb potential $W$ is calculated from the inverse of the macroscopic dielectric tensor, evaluated within the random-phase approximation. 
Details on the implementation within the LAPW formalism can be found in Refs.~\onlinecite{pusc-ambr02prb,sagm-ambr09pccp}.
The coefficients $\gamma_x$ and $\gamma_c$ in Eq.~\eqref{eq:H_BSE} enable \textit{turning on} and \textit{off} the exchange and the direct term, $\hat{H}^x$ and $\hat{H}^{dir}$ (Eq.~\ref{eq:Hx} and~\ref{eq:Hdir}), respectively.
The solutions of the full Hamiltonian in Eq.~\ref{eq:H_BSE} ($\gamma_x$ = $\gamma_c$ = 1) correspond to \textit{singlet} excitations.
When the $e$-$h$ exchange interaction is neglected ($\gamma_x$ = 0 and $\gamma_c$ = 1) \textit{triplet} solutions are obtained.
Further details on the BSE Hamiltonian and on its spin structure can be found in Refs.~\onlinecite{rohl-loui00prb,pusc-ambr02prb}. 
For bound excitons below the QP gap ($E_g$) obtained from $G_0W_0$ binding energies ($E_b$) are computed as the difference between $E_g$ and the excitation energies $E^{\lambda}$.
The eigenvectors $A^{\lambda}$ of Eq.~\ref{eq:H_BSE} provide information about the character and composition of the excitations.
In particular, they are used to define the weight of each transition between valence and conduction states at a given $\mathbf{k}$-point:
\begin{equation}
w^{\lambda}_{v\mathbf{k}} = \sum_c |A^{\lambda}_{vc\mathbf{k}}|^2 , \, \, \, w^{\lambda}_{c\mathbf{k}} = \sum_v |A^{\lambda}_{vc\mathbf{k}}|^2,
\label{eq:w}
\end{equation}
where the sums are performed over the range of occupied and unoccupied states included in the solution of the BSE (Eq.~\ref{eq:H_BSE}).
The eigenvectors $A^{\lambda}$ enter the expression of the oscillator strength, given by the square modulus of
\begin{equation}
\mathbf{t}^{\lambda}= \sum_{vc\mathbf{k}} A^{\lambda}_{vc\mathbf{k}} \dfrac{\langle v\mathbf{k}|\widehat{\mathbf{p}}|c\mathbf{k}\rangle}{\varepsilon_{c\mathbf{k}} - \varepsilon_{v\mathbf{k}}} ,
\label{eq:t}
\end{equation}
where $\hat{\mathbf{p}}$ is the momentum operator, and $\varepsilon_{v\mathbf{k}}$ and $\varepsilon_{c\mathbf{k}}$ are the QP energies of the involved valence and conduction states, respectively.
The optical absorption spectra are represented by the imaginary part of the macroscopic dielectric function ($\epsilon_M$)
\begin{equation}
\mathrm{Im}\epsilon_M = \dfrac{8\pi^2}{\Omega} \sum_{\lambda} |\mathbf{t}^{\lambda}|^2 \delta(\omega - E^{\lambda}) ,
\label{eq:ImeM}
\end{equation}
where $\Omega$ is the unit cell volume and $\omega$ is the energy of the incoming photon.
%%%%%%%%%%%%%%%%%%%%%
\section{Computational details}
All calculations are performed with the all-electron full-potential code \texttt{exciting}, \cite{gula+14jpcm} implementing the LAPW method.
The ground state of the investigated systems is computed by means of DFT, adopting the Perdew-Wang local-density approximation for the exchange-correlation functional. \cite{perd-wang92prb}
For the sampling of the Brilloiun zone (BZ), we employ a 6$\times$4$\times$1 (3$\times$2$\times$1) $\mathbf{k}$-point mesh for the p-SAM (d-SAM) in both ground-state and many-body perturbation theory calculations. 
Concerning the basis functions, a planewave cutoff $G_{max}$=5\,bohr is adopted for the calculation of the single molecule.
For the SAMs, this value is 4.625\,bohr.
Muffin-tin spheres with radii of 0.8\,bohr are considered for hydrogen, 1.1\,bohr for nitrogen, and 1.2 bohr for carbon and oxygen.
Each structure is optimized by minimizing atomic forces within a threshold of 0.025\,eV/\AA{}, with the lattice parameters of the unit cell kept fixed.
The corresponding geometries are displayed in Figs.~\ref{fig:H-az}b-d. 

In our $G_0W_0$ implementation, \cite{nabo+16prb} the dynamically screened Coulomb potential $W_0$ is computed within the random-phase approximation.
200 empty Kohn-Sham (KS) states are included in the calculation.
The so-obtained correction of 3.42\,eV, 3.32\,eV, and 1.8\,eV to the band gap of the isolated molecule, the d-SAM, and the p-SAM, respectively, is then applied to the KS electronic structure through a scissors operator.
BSE calculations are performed within the Tamm-Dancoff approximation.
To calculate the screened Coulomb interaction 500 unoccupied states for the molecule and 400 conduction bands for the SAMs are considered.
Approximately 1000 (500) $|\mathbf{G}+\mathbf{q}|$ vectors are included for the d-SAM (p-SAM) and more than 5000 for the single molecule.
In the solution of the BSE Hamiltonian (Eq.~\ref{eq:H_BSE}) for the single molecule we include 16 occupied and 22 unoccupied states, corresponding to about 8 and 5\,eV below and above the Fermi energy, respectively.
In the case of the d-SAMs we consider 11 valence and 15 conduction bands, corresponding to 5 and 4.5\,eV below and above the Fermi energy, respectively.
For the p-SAM, 10 occupied and 11 unoccupied bands are taken into account, corresponding to 3 and 4\,eV below and above the Fermi energy, respectively.

For a quantitative comparison with the experimental data, we calculate optical coefficients, specifically absorbance and reflectance, by adopting the 4$\times$4 matrix formulation of Maxwell's equations, \cite{yeh80ss,pusc-ambr06aem} as implemented in the \texttt{LayerOptics} code. \cite{vorw+16cpc}

%%%%%%%%%%%%%%%%%%%%%%%%%%%%%%%%%%%
%*******************************
%  RESULTS and DISCUSSION
%*******************************
\section{Results and Discussion}

\begin{figure*}
\centering
\includegraphics[width=.95\textwidth]{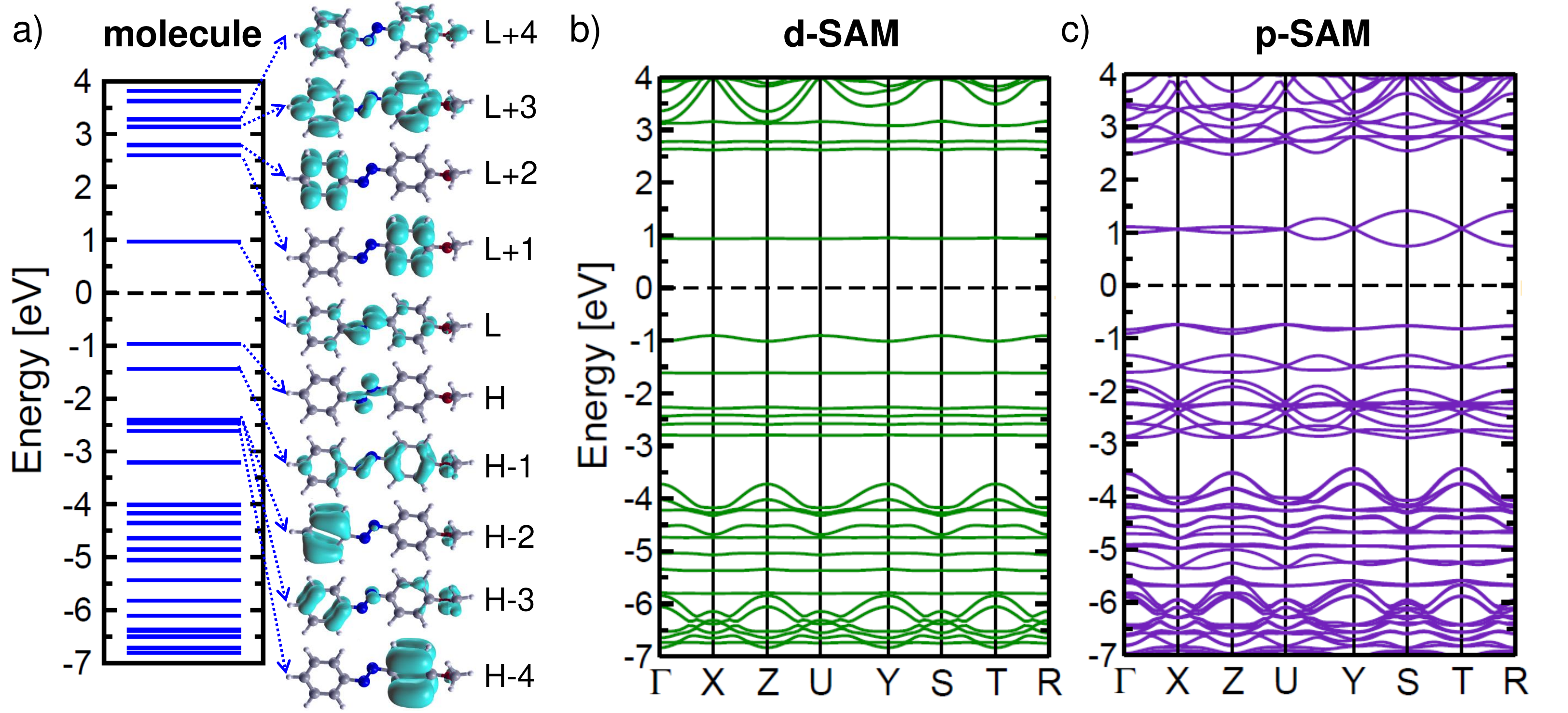}%
\caption{(Color online) a) Kohn-Sham energy levels of a single azobenzene molecule with the corresponding molecular orbitals; DFT band structures of b) d-SAM and c) p-SAM. The Fermi energy  in the mid gap is set to zero and indicated by a dashed line. 
}
\label{figure2_bands}
\end{figure*}
%

%%%%%%%%%%%%%%%%%%%%%%%%%%
\subsection{Electronic properties}
\label{sec:electronic}
%%%%%%%%%%%%%%%%%%%%%%%%%%
We start our analysis by inspecting the KS electronic properties of the isolated azobenzene and of the SAMs. 
In Fig.~\ref{figure2_bands} we show the energy levels of the single molecule, along with the real-space representation of a set of molecular orbitals closest to the Fermi energy.
In the proximity of the KS gap, we find a number of states with pronounced localization on the azo group.
These findings are in agreement with previous studies on \textit{trans}-azobenzene in the gas phase. \cite{crec-roit06jpca,cont+08jacs,maur-reut11jcp}
In particular, the highest occupied molecular orbital, HOMO (H), is distributed almost exclusively on the \ce{N=N} bond, thus being a non-bonding ($n$) orbital.
On the other hand, the H-1 and the lowest unoccupied molecular orbital, LUMO (L) exhibit $\pi$ character with the electron density spread over the phenyl rings.
Further away from the KS gap, the states are localized only on either carbon ring: this is the case of L+1 and L+2, as well as of H-2 and H-4. 
These pairs of states are energetically very close to each other.
In between, H-3 is spread over the whole molecule, including the \ce{OCH3} group.
In the conduction region, additional delocalized levels, such as L+3 and L+4, are found at higher energies, more than 3\,eV above the gap.

The electronic properties of the single molecule are reflected also in the SAMs.
By inspecting the band structure of the d-SAM (Fig.~\ref{figure2_bands}b), which includes only one azobenzene in the unit cell, we notice that the KS gap has basically the same size as in the isolated molecule.
The valence-band maximum (VBM) and the conduction-band minimum (CBM) maintain their localized character also in the SAMs.
In the mean-field DFT picture, their energy difference is almost unaffected by the packing arrangement. 
Also the very low dispersion of the bands near the fundamental gap confirms the localized molecular character of the electronic states in the d-SAM.
In the valence region, $\sim$2.5\,eV below the Fermi energy ($E_F$), a set of almost flat bands appears. 
Again, they are similar to the KS states of the isolated molecule: VBM-2 is rather localized on the \ce{OCH3} group, whereas VBM-3 and VBM-4 are the counterparts of H-2 and H-4, respectively.
The lowest-energy state in this manifold of bands has delocalized $\pi$ character.
Also in the conduction region, we notice strong similarities with the electronic properties of the isolated azobenzene.
About 1.5\,eV above the lowest unoccupied band, we find three flat bands, with CBM+1 and CBM+2 being the counterparts of L+1 and L+2, respectively.
The highest-energy band of this group has instead extended $\pi^*$ character, like the L+3 state in Fig.~\ref{figure2_bands}a.
Below -4\,eV and above 3\,eV, the KS states exhibit a delocalized intermolecular character, as confirmed by the more pronounced dispersion of the corresponding bands.

In Fig.~\ref{figure2_bands}c the band structure of the p-SAM is shown.
Since this system has two molecules per unit cell (Fig.~\ref{fig:H-az}c), bands have double multiplicity.
This is especially evident at the top of the valence region, where the two uppermost nearly dispersionless bands, VBM and VBM-1, have almost the same energy.
They are localized on the azo group and hence exhibit $n$ character like the HOMO in the isolated molecule.
At lower energy in the valence region we find another pair of bands with more pronounced dispersion, namely VBM-2 and VBM-3.
The latter are $\pi$ states, corresponding to H-1 in the single molecule (Fig.~\ref{figure2_bands}a).
Even deeper in energy, about 2\,eV below $E_F$, a manifold of bands appears, including states which are the counterparts of H-2, H-3, and H-4 of the isolated molecule.
In the conduction region, the two lowest-energy bands (CBM and CBM+1) are almost flat and degenerate along the $\Gamma$-X direction, which is almost parallel to the projection of the long molecular axis onto the two-dimensional plane of the SAM (see Fig~\ref{fig:H-az}c).
These states are delocalized over the whole molecule, similar to the LUMO of the isolated compound (Fig.~\ref{figure2_bands}a).
A gap of about 1.5\,eV separates these two lowest unoccupied states from higher-energy ones, which exhibit much more pronounced dispersion.
In this region the KS states show enhanced intermolecular coupling, which results in an increased wave-function delocalization.

\begin{figure}[h!]
\centering
\includegraphics[width=.48\textwidth]{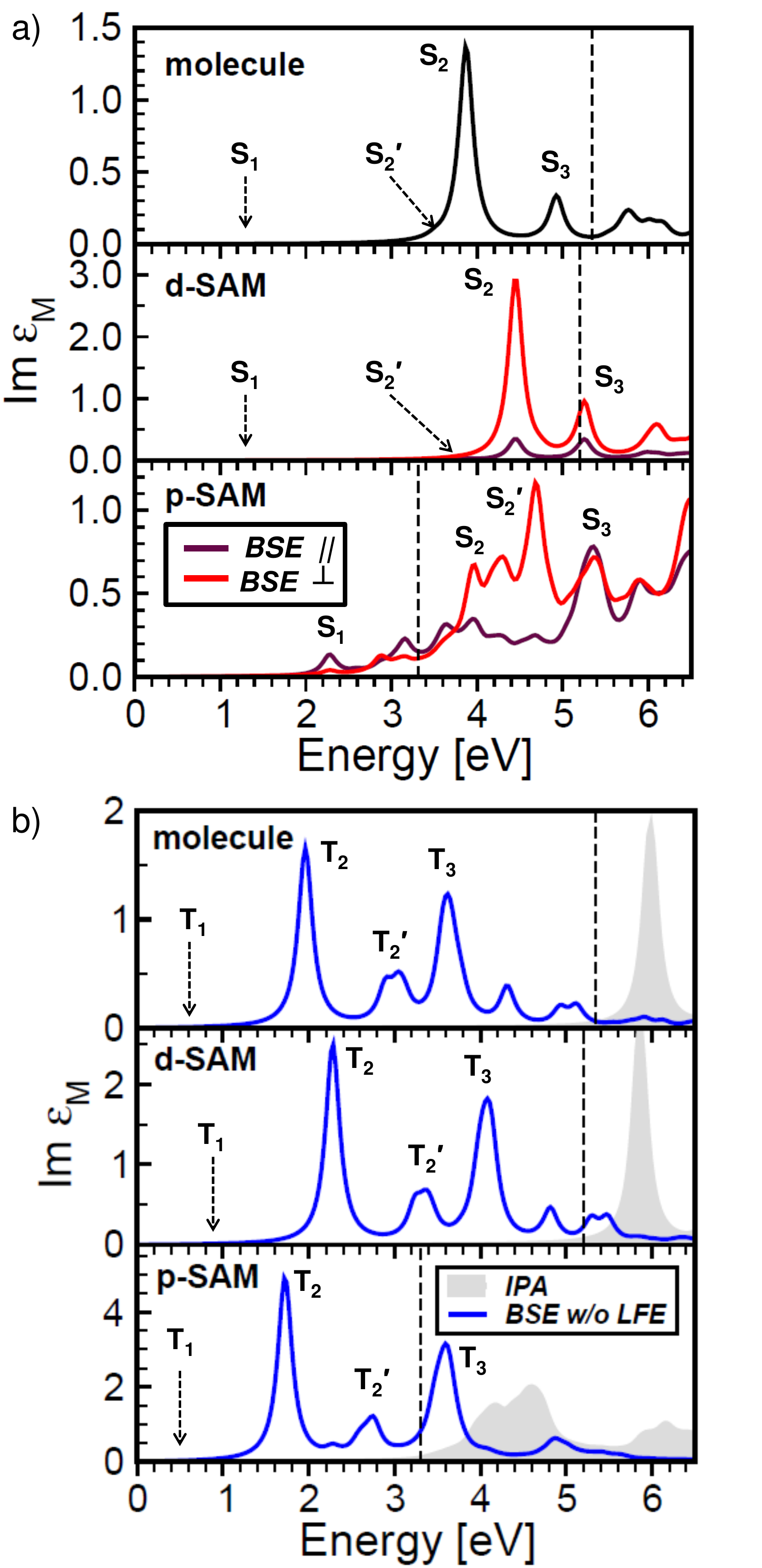}%
\caption{(Color online) Optical spectra of azobenzene molecule and SAMs, expressed by Im$\mathrm{\epsilon_M}$. The $G_0W_0$ gap is marked by a vertical dashed line. a) Singlet excitations, computed from the full BSE. An average over the three Cartesian components is shown for the molecule, while the in-plane ($\parallel$) and out-of-plane ($\perp$) components are plotted for the SAMs. b) BSE spectra, averaged over all Cartesian components, computed without LFE (\textit{triplet} excitations, solid line) and within the independent-particle approximation (IPA, shaded area). A Lorentzian broadening of 0.1\,eV is applied to all the spectra.
}
\label{figure3_BSE-spectra}
\end{figure}
%

%%%%%%%%%%%%%%%%%%%%%%%%%%
\subsection{Optical properties}
\label{sec:optics}
%%%%%%%%%%%%%%%%%%%%%%%%%%

%TABLE 1 - EXCITATION ENERGIES ALL
\begin{table}
\begin{tabular}{c||l|l||l|l||l}
& & \textbf{BSE} & & \textbf{BSE w/o LFE} & \textbf{IPA} \\ \hline \hline
% molecule
\multirow{4}{*}{\textbf{Molecule}} & \textbf{$S_1$} & 1.25 (\textit{4.09}) & \textbf{$T_1$} & 0.63 (\textit{4.71}) & 5.33   \\
 & \textbf{$S_2$} & 3.87 (\textit{1.47}) & \textbf{$T_2$} & 1.96 (\textit{3.38}) &  5.98 \\
 & \textbf{$S_2$'} & 3.50 (\textit{1.81}) & \textbf{$T_2$'} & 2.89 -- 3.11 (\textit{2.45}) & 6.80  \\
 & \textbf{$S_3$} & 4.93 (\textit{0.41}) & \textbf{$T_3$} & 3.56 -- 3.66 (\textit{1.78}) & 7.70 \\ \hline \hline
 % d-SAM
\multirow{4}{*}{\textbf{d-SAM}} & \textbf{$S_1$} & 1.29 (\textit{3.91}) & \textbf{$T_1$} & 0.88 (\textit{4.32})   & 5.20 \\
 & \textbf{$S_2$} & 4.45 (\textit{0.76}) & \textbf{$T_2$} & 2.28 (\textit{2.93}) & 5.85 \\
 & \textbf{$S_2$'} & 3.79 (\textit{1.42}) & \textbf{$T_2$'} & 3.24 (\textit{1.97}) &  6.64  \\
 & \textbf{$S_3$} & 5.25 & \textbf{$T_3$} & 4.05 (\textit{1.16}) & 7.58   \\ \hline \hline 
% p-SAM
\multirow{4}{*}{\textbf{p-SAM}} & \textbf{$S_1$} & 2.27 (\textit{1.03}) & \textbf{$T_1$} & 0.55 (\textit{2.76}) &   \\
 & \textbf{$S_2$} & 3.96 -- 4.01 & \textbf{$T_2$} & 1.72 (\textit{1.59}) &  \\
 & \textbf{$S_2$'} & 4.15 -- 4.30 & \textbf{$T_2$'} & 2.75 (\textit{0.56}) &   \\
 & \textbf{$S_3$} & 4.67 -- 4.70 & \textbf{$T_3$} & 3.54 &   \\ \hline \hline 
\end{tabular}
\caption{Excitation energies and binding energies (for bound excitons only, in parenthesis) of the \textit{singlet} (S) and \textit{triplet} (T) excitations marked in the spectra in Fig.~\ref{figure3_BSE-spectra}. In case of more than one excitation contributing to the peak, an energy range is listed. All energies are expressed in eV.}
\label{table1}
\end{table}

In Fig.~\ref{figure3_BSE-spectra}a, the calculated optical spectra of the azobenzene molecule and of the SAMs are shown. 
The energies of the main excitations are summarized in Table~\ref{table1}.
We start our analysis from the spectrum of an isolated molecule (Fig.~\ref{figure3_BSE-spectra}a, top panel), which is expressed by Im$\epsilon_M$ averaged over the three Cartesian components.
The lowest-energy excitation, $S_1$, is forbidden by symmetry.
It has $n$-$\pi^*$ character, corresponding to an almost pure transition between the HOMO and the LUMO, as reported in Table~\ref{table2}.
At higher energy, at about 3.9\,eV, a strong peak dominates the spectrum.
This excitation, labeled $S_2$, is responsible for the photo-isomerization of azobenzene, switching from the \textit{trans} to the \textit{cis} conformation. \cite{rau-lued82jacs}
$S_2$ is a bound exciton, with $E_b \sim$1.5\,eV.
It exhibits $\pi$-$\pi^*$ character, being dominated by the H-1$\rightarrow$L transition.
At 3.5\,eV $S_2$' appears as a weak shoulder of $S_2$, given by two almost degenerate excitations from H-2 and H-3 to the LUMO.
The localization of these occupied states on a single phenyl ring (Fig~\ref{figure2_bands}a) explains the low oscillator strength of $S_2$'.
The last bound exciton appearing in the spectrum is $S_3$, which shows a remarkably mixed character with contributions from a number of occupied $\pi$ states to the $\pi^*$ orbitals L+1 and L+2 (see Table~\ref{table2}).
The excitation energies of the molecule considered here are lower by a few hundred meV compared to those reported in the literature for bare \textit{trans}-azobenzene. \cite{catt-pers99pccp,astr+00jacs,flie+03jacs,cont+08jacs,maur-reut11jcp}
This is expected, since the presence of the oxygen-based group terminating the molecule (see Fig.~\ref{fig:H-az}a) is known to cause a red-shift of the $\pi$-$\pi^*$ transitions. \cite{utec+11pccp}
On the other hand, the excitation energy of $S_1$ is underestimated by more than 1.5\,eV compared to results from literature (see, \textit{e.g.}, Ref.~\onlinecite{crec-roit06jpca} and references therein). 
This can be ascribed to the starting-point dependence of the $G_0W_0$ step, which becomes particularly severe in case of localized KS states, such as the HOMO of azobenzene.
We expect that a hybrid functional as starting point for the $G_0W_0$ calculation \cite{brun-marq12jctc} or self-consistent $GW$ \cite{maro+12prb} would improve the result.
However, this problem does not affect the nature of the excitations, and therefore it is irrelevant for the essence of the present work.

% TABLE 2
\begin{table*}
\begin{tabular}{c|c|c|c|c}
 % singlet
  & \textbf{$S_1$} & \textbf{$S_2$} & \textbf{$S_2$'} & \textbf{$S_3$} \\ \hline \hline
 \multirow{4}{*}{\textbf{BSE singlet}} & \multirow{4}{*}{H$\rightarrow$L (98$\%$)} & \multirow{4}{*}{H-1$\rightarrow$L (89$\%$)} & H-3$\rightarrow$L (74$\%$) & H-1$\rightarrow$L+2 (22$\%$) \\
  & & & H-1$\rightarrow$L+2 (12$\%$) & H-3$\rightarrow$L+3 (17$\%$) \\ 
  & & & & H-1$\rightarrow$L+1 (15$\%$) \\ 
  & & & & H-3$\rightarrow$L (11$\%$) \\ \hline \hline
 %triplet
  & \textbf{$T_1$} & \textbf{$T_2$} & \textbf{$T_2$'} & \textbf{$T_3$} \\ \hline \hline
\multirow{5}{*}{\textbf{BSE triplet}} & \multirow{5}{*}{H$\rightarrow$L (95$\%$)} & \multirow{5}{*}{H-1$\rightarrow$L (94$\%$)} & H-3$\rightarrow$L (81$\%$) & H-3$\rightarrow$L+2 (79$\%$) \\ \cline{4-5}
 & & & H-4$\rightarrow$L (32$\%$) & H-1$\rightarrow$L+1 (72$\%$) \\ 
 & & & H-5$\rightarrow$L (41$\%$) & H-5$\rightarrow$L+1 (15$\%$) \\ \cline{4-5}
 & & & & H-4$\rightarrow$L+1 (76$\%$) \\ 
 & & & & H-2$\rightarrow$L+1 (12$\%$) \\ \hline \hline
% IPA
\multirow{2}{*}{\textbf{IPA}} & \multirow{2}{*}{H$\rightarrow$L} & \multirow{2}{*}{H-1$\rightarrow$L} & H-3$\rightarrow$L & H-1$\rightarrow$L+1 \\ \cline{4-5}
 & & & H-4$\rightarrow$L & H-1$\rightarrow$L+2 \\ \hline \hline
\end{tabular}
\caption{Composition, in terms of single-particle transitions, of the main excitations of the isolated azobenzene molecule. The weight of each transition is given in brackets, including only contributions larger than 10$\%$. In BSE-\textit{triplet} and IPA the contributions of all active excitations embraced by the peaks are listed.}
\label{table2}
\end{table*}

For the analysis of the absorption spectra of the SAMs, we plot the parallel ($\parallel$) and perpendicular ($\perp$) components of the imaginary part of the macroscopic dielectric function: Im$\epsilon_M^{\parallel}$ represents an average of the $ab$ plane of the unit cell depicted in Fig.~\ref{fig:H-az}c-d, while Im$\epsilon_M^{\perp}$ corresponds to the component along the $c$ axis.
The overall pronounced differences reflect the orientation of the molecules in the SAM.
The spectrum of the d-SAM exhibits strong similarities with the one of the single molecule, starting from the position of the QP gap.
In the lowest energy region, we find the n-$\pi^*$ excitation $S_1$, which is still symmetry-forbidden.
At higher energy ($\sim$4.5\,eV), the strong $\pi$-$\pi^*$ peak $S_2$ appears.
Its excitation energy is blue-shifted by about 0.6\,eV compared to the isolated molecule, as a result of two counteracting effects.
On the one hand, the QP gap, represented by the IPA onset (Table~\ref{table1} and Fig.~\ref{figure3_BSE-spectra}b) decreases by 0.13\,eV in the d-SAM, due to intermolecular coupling.
On the other hand, the exciton binding energy is reduced by almost 0.4\,eV, owing to the enhanced screening and wave-function overlap resulting from the packing of the chromophores in the unit cell.
A similar effect is noticed also for $S_3$, which is unbound, \textit{i.e.}, appears above the QP gap.
Compared to the spectrum of the isolated molecule, the excitation energy of $S_3$ undergoes a blue-shift of more than 0.3\,eV (Table~\ref{table1}).

\begin{figure*}
\centering
\includegraphics[width=.98\textwidth]{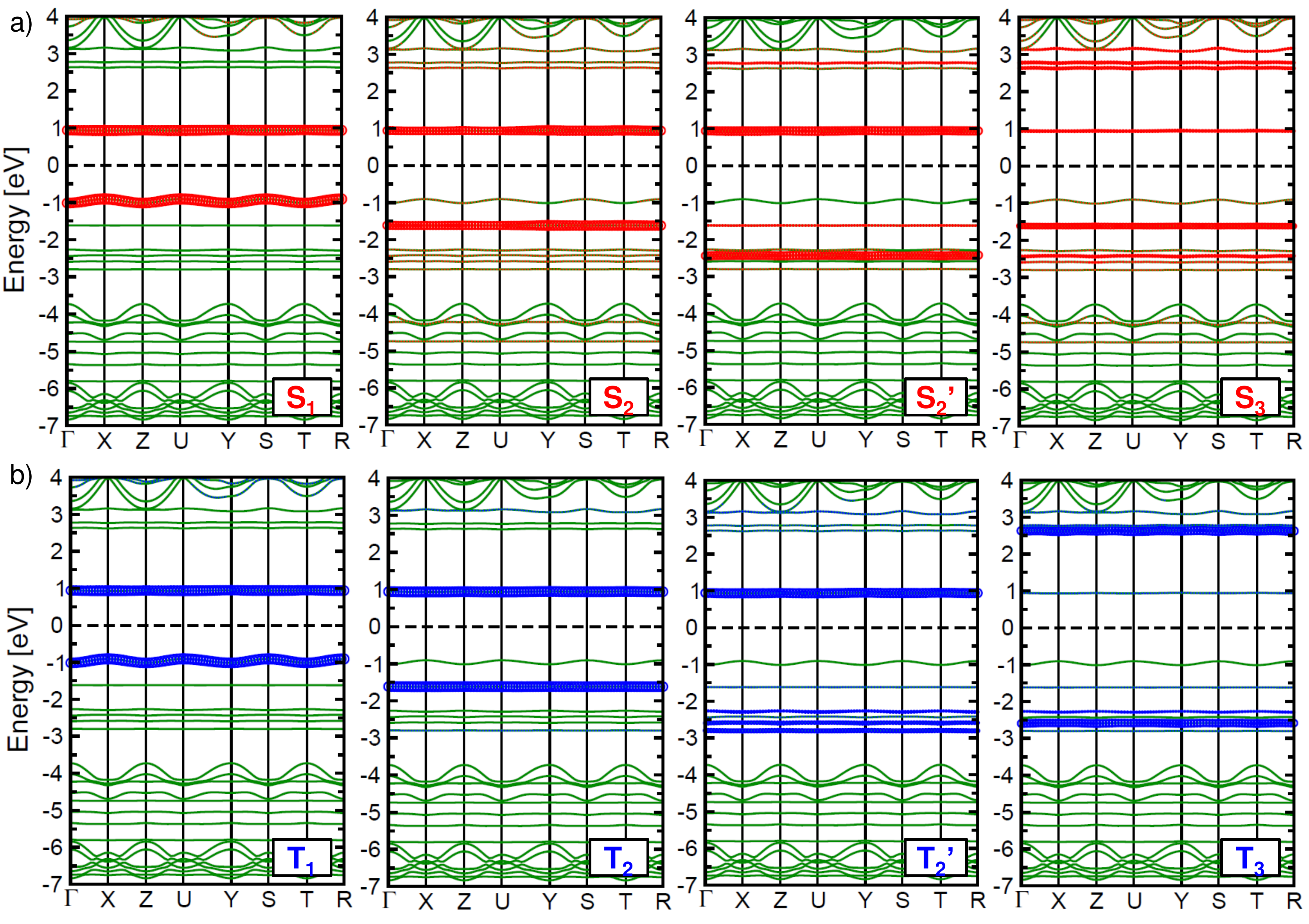}%
\caption{(Color online) Weights of \textit{singlet} (a) and \textit{triplet} (b) excitations of the d-SAM plotted on top of the Kohn-Sham band structure. The size of the red and blue circles is representative of their magnitude.
}
\label{excitons_d-SAM}
\end{figure*}

For a quantitative analysis of the character of these excitations, we display in Fig.~\ref{excitons_d-SAM} the corresponding weights (Eq.~\ref{eq:w}) on top of the KS band structure.
Overall, the excitations of the d-SAM show analogous character to those of the isolated molecule.
$S_1$ is an almost pure transition between the VBM and the CBM, which are the counterparts of the HOMO and the LUMO, respectively.
In the same way, $S_2$ is clearly dominated by the transition from the VBM-1 to the CBM, and $S_2$' by VBM-3 to CBM.
Also in the d-SAM, $S_3$ exhibits a remarkably mixed character, with the most significant contributions coming from  VBM-1 and VBM-3 in the valence region and from the four lowest conduction bands.
The polarization of $S_2$ and $S_3$ along the long molecular axis is further emphasized by the predominance of Im$\epsilon_M^{\perp}$ over Im$\epsilon_M^{\parallel}$ (Fig.~\ref{figure3_BSE-spectra}a).

\begin{figure*}
\centering
\includegraphics[width=.98\textwidth]{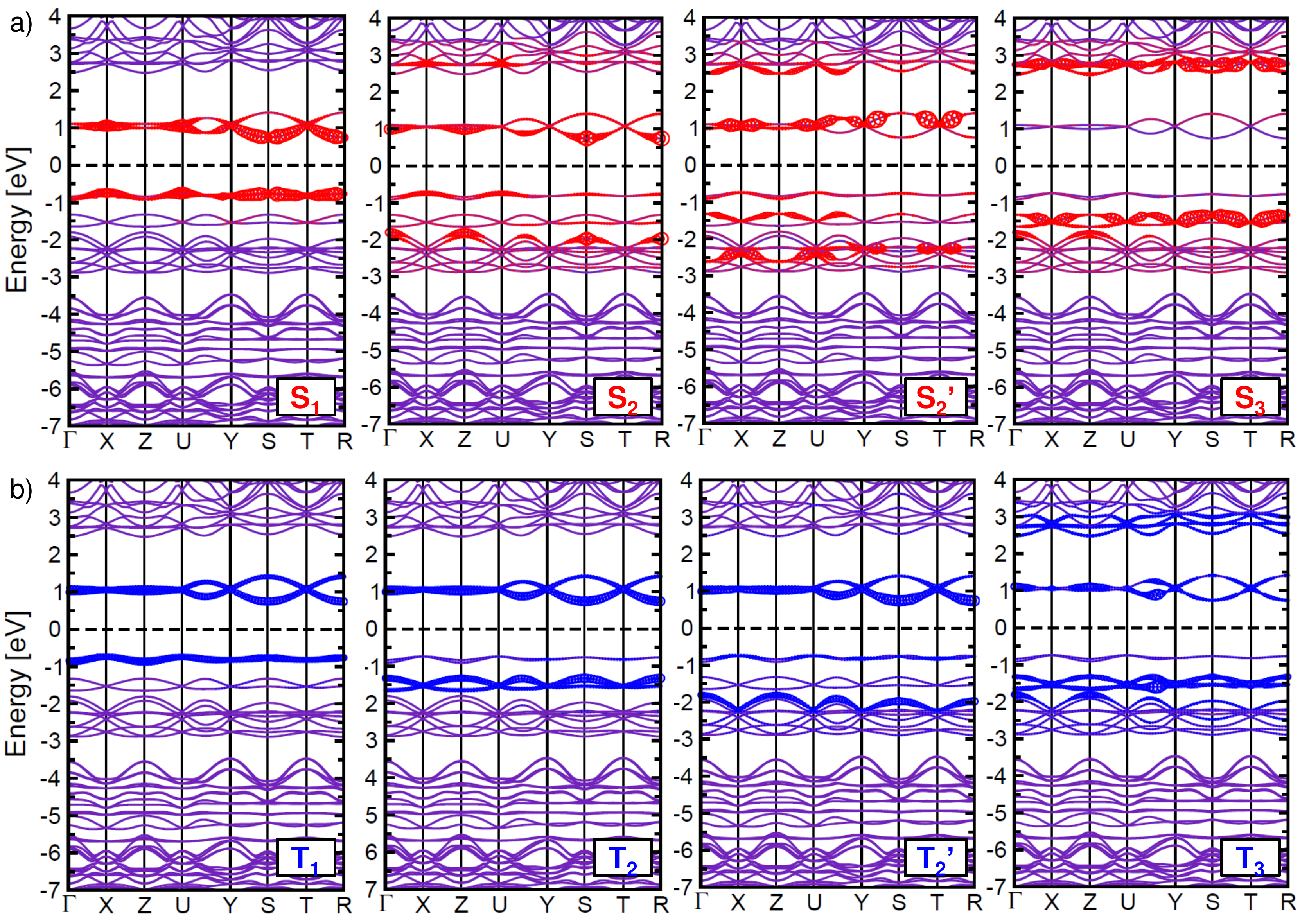}%
\caption{(Color online) Weights of \textit{singlet} (a) and \textit{triplet} (b) excitations of the p-SAM plotted on top of the Kohn-Sham band structure. The size of the red and blue circles is representative of their magnitude.
}
\label{excitons_p-SAM}
\end{figure*}

The optical spectrum of the p-SAM is completely different from the previous ones.
With increasing packing density of the chromophores and thus enhanced intermolecular interactions, the optical gap, represented by the IPA absorption onset, is significantly reduced to about 3.5\,eV (Table~\ref{table1}).
In addition, two specific features are immediately visible in the spectrum displayed on the bottom panel of Fig.~\ref{figure3_BSE-spectra}a.
First, the lowest-energy excitation $S_1$ is optically active, although weak.
It is still found below the QP gap, with $E_b \approx$~1\,eV.
Second, the sharp peak $S_2$, associated with a bound excitonic state in the spectra of the molecule and of the d-SAM, now appears above the absorption onset of the material in a broad absorption band along with other intense excitations.
This behavior is mainly due to LFE, counteracting the red-shift induced by the band-gap narrowing, as an effect of intermolecular electronic interactions. \cite{hahn+05prl,herm+08prl,herm-schw11prl}
Moreover, contrary to the case of the d-SAM where the out-of-plane component of Im$\epsilon_M$ dominates the spectrum in the considered frequency range, here Im$\epsilon_M^{\parallel}$ and Im$\epsilon_M^{\perp}$ show different spectral weights in specific energy regions. 
This is obviously related to the character of the excitations involved.
To gain deeper insight, we again consider the excitation weights plotted on top of the KS band structure (Fig.~\ref{excitons_p-SAM}).
As discussed in Sec.~\ref{sec:electronic}, the presence of two molecules in the unit cell gives rise to two bands at the top of the valence region (VBM and VBM-1) and at the bottom of the conduction region (CBM and CBM+1) having the same character as the HOMO and the LUMO of the isolated molecule.
The first exciton $S_1$ arises from a transition between these occupied and unoccupied states, and is now dipole-allowed, due to symmetry breaking.
The weights are homogeneously distributed within the BZ in both valence and conduction regions revealing the intramolecular character of this excitation.
The $n$-$\pi^*$ nature of $S_1$ explains the predominance of Im$\epsilon_M^{\parallel}$ at the absorption onset (Fig~\ref{figure3_BSE-spectra}a).
Around 3\,eV, where a manifold of excitations with similar character as $S_1$ takes place, the in-plane component of the macroscopic dielectric function dominates over the out-of-plane one.
Excitations with $\pi$-$\pi^*$ character start appearing $\sim$1\,eV above the QP gap, at about 4\,eV, where also the relative intensity of Im$\epsilon_M^{\perp}$ overcomes the one of Im$\epsilon_M^{\parallel}$.
In the spectrum of the p-SAM, we identify $S_2$ as the most intense excitation among a manifold of solutions of the BSE exhibiting rather mixed character. 
From Fig.~\ref{excitons_p-SAM}a we notice that the largest weights come from transitions between VBM-5 and VBM-6 to CBM and CBM+1, and non-negligible contributions arise also from transitions from the VBM to higher conduction bands.
The weights are not uniformly distributed throughout the BZ, pointing to an intermolecular character of the excitation. 
Excitations with the same mixed character but larger contributions from transitions between VBM-2/VBM-3 to CBM/CBM+1 appear in the same energy window (3.9 -- 4.5\,eV).
However, they display lower oscillator strength than $S_2$.
Also $S_2$' shows a very mixed character, and is dominated by contributions similar to those of $S_2$.
Again, we notice an inhomogeneous distribution of the weights in the BZ, indicating an intermolecular character of the $e$-$h$ wave-function in real space.
Finally, we identify $S_3$, in analogy with the molecule and the d-SAM, as an excitation from VBM-2 and VBM-3 to higher-energy conduction bands, namely CBM+2 and CBM+3.

The striking differences in the optical absorption properties of the p-SAM compared to its diluted counterpart and the isolated molecule are mainly to be ascribed to intermolecular interactions that act in different ways.
The array of closely-packed molecules gives rise to enhanced screening effects, related to more delocalized $e$-$h$ wave-functions, which exhibit a reduced binding strength.
From the spectra in Fig.~\ref{figure3_BSE-spectra}a and the data reported in Table~\ref{table1}, this trend is very clear.
Considering for example $S_2$, we notice that its binding energy decreases systematically with increasing molecular concentration, up to the p-SAM, where this excitation is found more than 0.5\,eV above the QP gap.
Remarkably, a qualitatively similar behavior is exhibited by the core-level excitations from the nitrogen $K$-edge of azobenzene-functionalized SAMs. \cite{cocc-drax15prb}
In that case, the binding energy of the first bound exciton is reduced by about 2\,eV going from the isolated molecule to the p-SAM.
Another important effect of intermolecular interactions is the redistribution of the spectral weight to higher energies, due to dipole coupling.
This mechanism has been rationalized also for azobenzene dimers from a quantum-chemistry perspective. \cite{utec+11pccp,bena-corn13jpcc}
In the present work we demonstrate that such spectral blue-shift is directly associated to LFE, expressed by the short-range $e$-$h$ exchange interaction (Eq.~\ref{eq:Hx}). 

In order to quantitatively determine the role of LFE in the spectra of azobenzene-functionalized SAMs, we now compare Fig.~\ref{figure3_BSE-spectra}a and Fig.~\ref{figure3_BSE-spectra}b.
In the latter, the dielectric function is calculated by neglecting LFE, \textit{i.e.}, by setting $\gamma_{x}$=0 in Eq.~\ref{eq:H_BSE}, thereby turning off the \textit{exchange} interaction.
The solutions of the resulting Hamiltonian are the \textit{triplet} excitations. 
Being spin-forbidden, they cannot be probed by an optical-absorption experiment.
However, the comparison between \textit{singlet} and \textit{triplet} spectra helps us explaining the role of local fields and excitonic effects in the optical excitations of these systems.
All spectra in Fig.~\ref{figure3_BSE-spectra}b show the same features, namely the lowest-energy dark excitation $T_1$, the intense peaks $T_2$ and $T_3$, and the weaker $T_2$' in between. 
The corresponding excitation energies are obviously red-shifted compared to their \textit{singlet} counterparts (Fig.~\ref{figure3_BSE-spectra}a), due to the missing $e$-$h$ repulsive term.
All the \textit{triplet} excitations exhibit similar relative intensity regardless of the packing density, and likewise their nature remains the same from the isolated monomer to the p-SAM.
Specifically, we inspect the character of the $\pi$-$\pi^*$ transition $T_2$ and compare it with its \textit{singlet} counterpart $S_2$.
When intermolecular interactions are negligible or missing, as in the case of the single molecule, this excitation stems almost completely from H-1$\rightarrow$ L, no matter whether LFE are included or not (Table~\ref{table2}).
In fact, already in the IPA spectrum of the molecule shown in Fig.~\ref{figure3_BSE-spectra}b, a sharp peak appears, given by the $\pi$-$\pi^*$ excitation $T_2$.
Also for the d-SAM we find an intense resonance at the onset of the IPA spectrum.
By comparing the character of $S_2$ and $T_2$ in this system (Fig.~\ref{excitons_d-SAM}), the growing influence of intermolecular interactions is evident.
While $T_2$ is given by a pure VBM-1$\rightarrow$ CBM transition, in $S_2$ we find minor contributions also from deeper (higher) valence (conduction) bands.
Also $S_3$ and $T_3$ exhibit a similar behavior.
While the mixed character of this excitation is enhanced upon inclusion of the Coulomb interaction already in the isolated molecule (Table~\ref{table2}), in the d-SAM we notice that many more interband transitions are involved in the composition of $S_3$ compared to $T_3$.

In case of the p-SAM, intermolecular interactions captured by LFE significantly affect the character of the bright excitations $S_2$, $S_2$', and $S_3$.
In the IPA spectrum, that is shown for comparison, the absorption onset is rather featureless.
In the BSE spectrum computed without accounting for LFE (\textit{triplet} excitations) excitonic peaks are prominent, and resemble in energy and nature those characterizing the spectra of the molecule and of the d-SAM.
$T_2$ is an almost pure transition from VBM-2 and VBM-3 to CBM and CBM+1.
Likewise, $T_2$' shows a predominant contribution from lower-energy valence bands to CBM/CBM+1.
$T_3$ retains its mixed character, already discussed for the molecule and for the d-SAM.
Concerning the excitation energies, we notice a similar behavior as in the \textit{singlet} spectra in Fig.~\ref{figure3_BSE-spectra}a: $T_2$ blue-shifts by more than 0.3\,eV from the molecule to the d-SAM.
It results from narrowing of the QP gap ($\sim$0.1\,eV), due to the increased molecule-molecule electronic interactions, and the enhanced exciton delocalization, which reduces the binding energy by about 0.5\,eV.
When going from the diluted to the packed SAM, the excitations undergo a red-shift by approximately 0.6\,eV.
In this case, the reduction of the QP gap is so large ($\sim$2\,eV) that it overcomes the pronounced decrease of exciton binding energies ($\sim$1.4\,eV, see Table~\ref{table1}).
It is worth noting that these spectral shifts are unrelated to LFE, which are not present in the calculation of \textit{triplet} excitations. 
Similar effects, due to the coupling of the electronic wave-functions and to the consequent delocalization of the resulting $e$-$h$ pairs as an effect of the increased intermolecular interactions, have been discussed already in the context of organic crystals. \cite{ruin+02prl,humm+04prl,hahn+05prl}
When the effects of local fields are accounted for, as in the \textit{singlet} spectra, they additionally contribute to an overall blue-shift of the oscillator strength and, most importantly, to its redistribution to higher energies.

At this point, we briefly discuss our results in the context of the existing theoretical literature. \cite{utec+11pccp,bena-corn13jpcc}
In the present work, we investigate optical absorption properties of azobenzene-functionalized SAMs from a solid-state physics perspective.
We treat the ordered SAMs as periodic systems, such that the electronic wave-functions are allowed to spread over the infinitely extended structure. 
In this way, the interactions between the chromophores are quantitatively taken into account in the ground state as well as in the excited state.
Moreover, our approach for calculating optical excitations based on the solution of the BSE enables us to consider explicitly the effects of the repulsive exchange and of the attractive direct $e$-$h$ Coulomb interaction, which counteract each other.
From the results of this analysis we assert that the blue-shift experienced by the main absorption peaks in the spectrum of the p-SAM is mainly due to intermolecular coupling.
LFE enhance the mixing of single-particle transitions contributing to the excitations, thereby turning a sharp resonance into a broad absorption band.
The $e$-$h$ pairs become delocalized, assuming intermolecular character.
In this regard, our findings are in line with the rationale expressed by previous theoretical works, \cite{utec+11pccp,bena-corn13jpcc} where SAMs have been modeled in a quantum-chemical framework, accounting for dipole-dipole interactions in molecular dimers and/or clusters. 

%%%%%%%%%%%%%%%%%%%%%%%%%%
\subsection{Differential reflectance spectra}
\label{sec:DRS}

\begin{figure*}
\centering
\includegraphics[width=.8\textwidth]{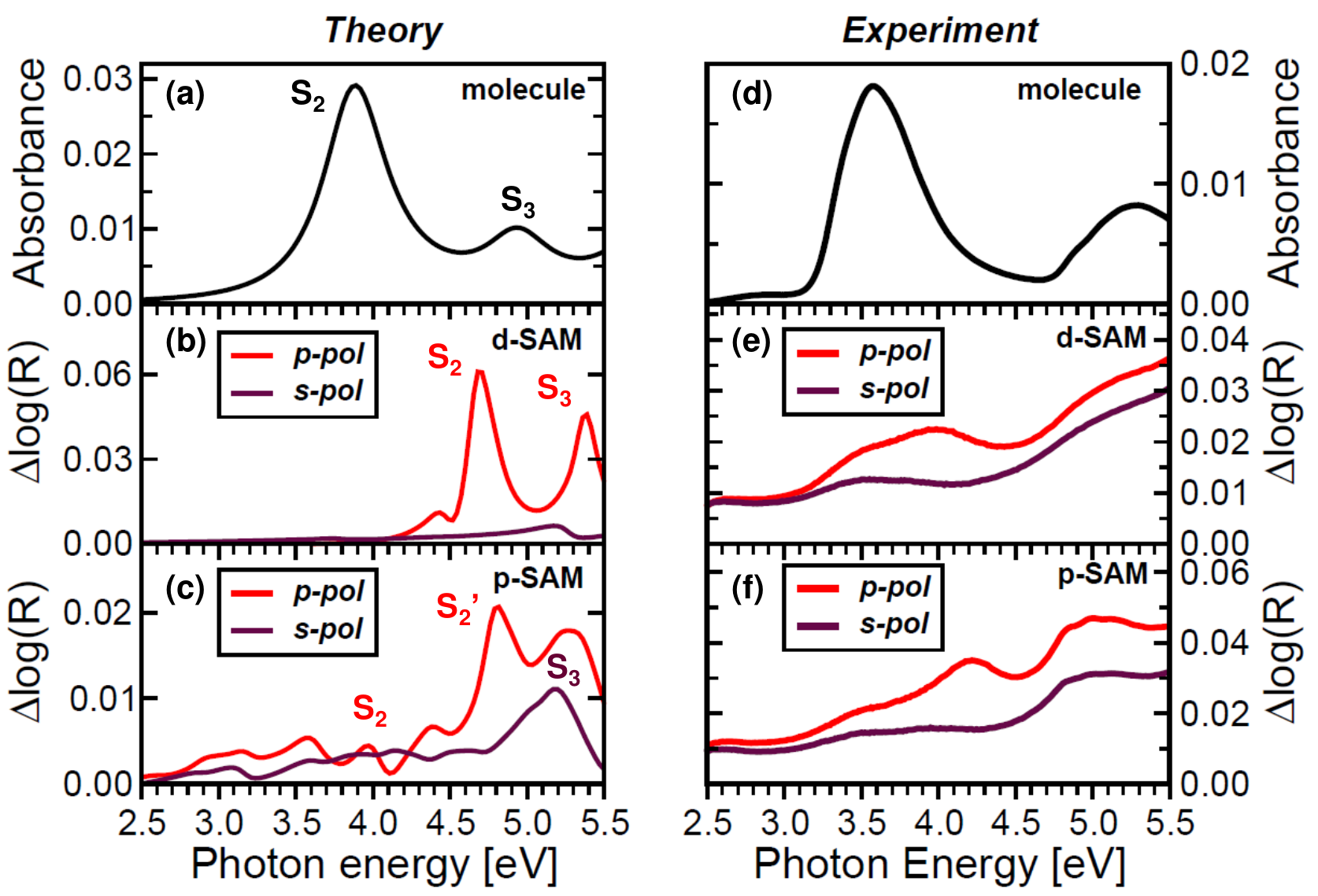}%
\caption{(Color online) Optical coefficients of the molecule and SAMs from theory (a-c) and experiment (d-f). The differential reflectance spectra of the SAMs with $s$- and $p$-polarized light are obtained under an angle of incidence of 45$^{\circ}$. A Lorentzian broadening of 0.25\,eV and 0.1\,eV is applied to the theoretical spectrum of the molecule (a) and of the SAMs (b,\,c), respectively.
}
\label{DeltaR}
\end{figure*}

To experimentally investigate the effect of increasing density of chromophores on the optical properties of azobenzene-functionalized alkanethiolate SAMs (Fig.~\ref{fig:H-az}a), we use the azo compound 11-(4-(phenyldiazenyl)phenoxy)-undecane-1-thiol (referred to as Az11) synthesized in the group of Rafal Klajn (Weizmann Institute of Science, Revohot, Israel), as well as dodecane-1-thiol (C12, Sigma-Aldrich).
A SAM is prepared by immersing a gold substrate into a methanolic solution of the thiols for 20 hours, as detailed in Ref.~\onlinecite{mold+15lang}. 
Depending on the relative concentrations of Az11 and C12 in solution, SAMs with different chromphore densities are obtained. 
Since the mixing of the molecules in the SAM is mainly statistical, the local environment of the azobenzene moieties is not homogeneous, in contrast to our first-principles calculations.
The optical properties of the SAM are determined by differential reflectance (DR) spectroscopy,  measuring the reflectance of the gold before and after the SAM preparation. 
All measurements are performed at ambient conditions, with an angle of incidence of 45$^{\circ}$, with both $s$- and $p$-polarized light. 
Additional experimental details are reported in Ref.~\onlinecite{mold+15lang}. 
For a direct comparison between theoretical and experimental data, we calculate DR spectra of azobenzene-functionalized SAMs on gold by employing the \texttt{LayerOptics} code. \cite{vorw+16cpc}
The contribution of the SAMs, assumed to form a 2 nm thick layer, is given by the full dielectric tensors computed from BSE.
The effects of the gold substrate are incorporated in a dielectric constant $\epsilon_{sub} = \epsilon_1 + i\epsilon_2$= -0.36 + $i$6.475.
This value is extracted from the average of the complex refractive index of gold, in the frequency region between 3 and 4.5\,eV, \cite{palik98book} corresponding to the $S_2$ excitation band of the azobenzene-functionalized SAM.
The choice of the dielectric permittivity of the metal substrate is crucial to capture the essential features of the experimental DR spectra.
The angle of incidence is set to 45$^{\circ}$ as in the experiment.
Interference effects at the layer boundaries, as well as the influence of the polarization of the incoming light are taken into account. 

In the case of the isolated molecule, absorption measurements are performed in solution.
The resulting absorbance curve is shown in Fig.~\ref{DeltaR}d, while its theoretical counterpart is displayed in Fig.~\ref{DeltaR}a.
Both spectra exhibit the same structure, with an intense peak at lower energy ($S_2$), followed by a weaker one at higher energy ($S_3$).
The first resonance $S_2$ is blue-shifted by approximately 0.3\,eV in the computed absorbance, compared to the experimental result. 
This difference can be ascribed to the solvent effects.
The relative intensity of the peaks $S_2$ and $S_3$ is well reproduced by theory, while the absolute energy position of the latter peak is underestimated by about 0.3\,eV.
This behavior can be attributed to the inclusion of the QP correction to the gap, associated to the self-energy of the LUMO, as a scissors operator, that rigidly up-shifts all conduction states by the same amount of energy. 
Considering that $S_3$ mostly stems from transitions to states above the LUMO (see Table~\ref{table2}), the application of a scissors operator can be responsible for the underestimation of its excitation energy.
Also the $G_0W_0$ starting-point, as discussed in Sec.~\ref{sec:optics}, can play a role in this regard.

The DR spectra of the packed and diluted SAMs are displayed in Fig.~\ref{DeltaR}\,b,\,c (theory) and Fig.~\ref{DeltaR}\,e,\,f (experiment).
In the first-principles results, the peaks are slightly shifted compared to the maxima in Im$\epsilon_M$ (Fig.~\ref{figure3_BSE-spectra}a).
Specifically, in the case of the d-SAM the first intense peak in Fig.~\ref{DeltaR}b is 0.25\,eV higher in energy with respect to the $S_2$ resonance identified in the macroscopic dielectric function in Fig.~\ref{figure3_BSE-spectra}a.
This difference is not unexpected, since the DR is not only obtained from the imaginary part of $\epsilon_M$, but also depends on its real part.
In agreement with the predominantly upright orientation of the chromophores, the $s$-polarized component is weak in the entire spectral window explored in Fig.~\ref{DeltaR}b, with a single broad peak at the low-energy edge of the $S_3$ band.
For the p-SAM, the energy region between 4.5 and 5.5\,eV is characterized by two peaks associated with $S_2$' ($p$-polarized component only) and $S_3$ (both $s$- and $p$-polarized components) in Im$\epsilon_M$ (Fig.~\ref{figure3_BSE-spectra}a).
At lower energy, around 4\,eV, we identify the spectral features associated with the $S_2$ resonance. 
Overall, the $s$-polarized component is enhanced in comparison with its counterpart in the d-SAM.
Between 3 and 4\,eV we notice a weak hump that can be related to a corresponding local maximum in the in-plane component of Im$\epsilon_M$ in Fig.~\ref{figure3_BSE-spectra}a.

The experimental DR spectra of the diluted and densely-packed SAMs (Fig.~\ref{DeltaR}\,e,\,f) are quite similar.
They both exhibit a broad hump in the region of the $S_2$ band of the isolated azobenzene, between approximately 3.2 and 4.5\,eV.
In the case of the d-SAM the maximum in the $p$-polarized component is shifted to higher energies by 0.4\,eV compared to the single molecule, while for the p-SAM this shift increases to 0.65\,eV.
It should be noted, however, that the peak maximum in the DR does not simply correspond to a blue-shifted $S_2$ resonance, but it rather originates from the multitude of excitations appearing in that energy region, as discussed in Sec.~\ref{sec:optics}.
The discrepancy between the measured spectrum of an Az11-SAM diluted with C12 chains and its theoretical counterpart (d-SAM) can be rationalized from the structural difference between the experimental sample and the system modeled from first principles.
While in the latter the orientation of the chromophores is assumed to be the same as in the p-SAM, the diluted SAM sample comprises a mixture of different local molecular environments.
The probability of finding two chromophores on directly neighboring sites is still very high for a dilution to 50$\%$. 
Additionally, the chromophores tend to orient themselves more parallel to the surface.
A detailed characterization of the diluted SAM samples is provided in Ref.~\onlinecite{mold+15lang}.
Since the interaction with nearest-neighbors contributes dominantly to the spectral shifts, the experimental DR spectra of diluted and densely-packed SAMs have a similar shape. 
Only the energetic position of the peak maximum changes with the chromophore density.

The trends shown by our theoretical and experimental results agree qualitatively, confirming the validity of our analysis and supporting the interpretation of our results.
As demonstrated by the analysis of our BSE spectra, local-fields enhance the effects of the Coulomb screening, giving rise to a significant blue-shift of the oscillator strength. 
As such, they play a crucial role in determining the spectral features at large packing density of the chromophores, being responsible for the pronounced mixing of single-particle transitions, which gives rise to the multitude of excitations in the spectral range of interest. 
Fine details in the spectra cannot be captured due to the differences between the first-principles description and the experimental setup discussed above. 
These differences, however, do not affect the essence of our results.

%****************************
%  CONCLUSIONS
%****************************
\section{Summary and Conclusions}
In summary, we have presented a joint theoretical and experimental study on the optical properties of azobenzene-functionalized SAMs. 
Our results indicate that at low molecular concentrations the spectra are dominated by an intense $\pi$-$\pi^*$ resonance in the near-UV region, which is known to be involved in the \textit{trans-cis} photo-isomerization of the azobenzene moiety.
At increasing packing density, this resonance is quenched, and the entire spectra undergo a significant redistribution of oscillator strengths to higher energies.
From a thorough analysis based on many-body perturbation theory, we are able to ascribe this behavior to a competition between band-gap narrowing, which tends to red-shift the spectra, and exciton delocalization, which blue-shifts the excitation energies, acting in the same direction as local-field effects.
The latter, in particular, are significantly enhanced by the strong coupling between the chromophores and are responsible for a remarkable mixing of the interband transitions contributing to the main excitations.
While clearly excitonic in nature, excitations become delocalized and the resulting absorption band is significantly broadened.
By lowering the concentration of molecules, the role of intermolecular coupling decreases accordingly, with the optical features of the isolated molecules being restored.

With this analysis we have revealed the fundamental physical mechanisms ruling the photo-absorption properties of azobenzene-functionalized SAMs.
The behavior of these systems upon light absorption is intrinsically dominated by many-body interactions. 
This knowledge represents an important step forward in view of understanding and rationalizing the excited-state properties of these complex systems.

\vspace{0.5cm}
Relevant input and output files of first-principles calculations can be found in the NoMaD Repository, with the corresponding DOI: http://dx.doi.org/10.17172/NOMAD/2016.12.07-1.

%%%%%%%%%%%%%%%%%%%%%%%%%%%%%%%%%%%%%%%%%%%%%%%%%%
\section*{Acknowledgement}
This work was funded by the German Research Foundation (DFG), through the Collaborative Research Center SFB 658. 
C.C. acknowledges financial support from the \textit{Berliner Chancengleichheitsprogramm} (BCP) and from IRIS Adlershof.
T.M., C.G., and M.W. thank Rafal Klajn and coworkers (Weizmann Institute of Science, Israel) for providing the azobenzene compound.

%%%%%%%%%%%%%%%%%%%%%%%%%%%%%%%%%%%%%%%%%%%%%%%%%%%
%\bibliography{bib}
%%%%%%%%%%%%%%%%%%%%%%%%%%%%%%

%%%%%%%%%%%%%%%%%%%%%%%%%%%%%%%%%%%%%%%%%%%%%%%%%%%%%%%%%%%%%%%%%%%%%

\end{document}